\documentclass[epsf,twocolumn,preprintnumbers]{revtex4}
\usepackage{graphics}
\usepackage{graphicx}
\usepackage{dcolumn} 
\usepackage{bm}
\usepackage{color}
\usepackage{epsfig}
\newcommand{\bni}{Ba$_2$NiO$_2$(AgSe)$_2$}

\newcommand{\sni}{Sr$_2$NiO$_2$(AgSe)$_2$}

\begin{document}
\title{Proposed ordering of textured spin singlets in a 
   bulk infinite layer nickelate}
\author{Hyo-Sun Jin$^1$}
\author{Warren E. Pickett$^{2}$}
\email{wepickett@ucdavis.edu}
\author{Kwan-Woo Lee$^{1,3}$}
\email{mckwan@korea.ac.kr}
\affiliation{
 $^1$Division of Display and Semiconductor Physics, Korea University, Sejong 30019, Korea\\
 $^2$Department of Physics, University of California, Davis, CA 95616, USA\\
 $^3$Department of Applied Physics, Graduate School, Korea University, Sejong 30019, Korea
}
\date{\today}
\begin{abstract}
The infinite-layer structure nickelate \bni~(BNOAS) with $d^8$ Ni ions and a peculiar
susceptibility $\chi(T)$, synthesized at high pressure, is studied with 
correlated density functional methods. The overriding feature of the calculations
is violation of Hund's rule coupled with complete but {\it unconventional 
spin-orbital polarization}, leading to an unexpected low spin $^1B_1$, 
``off-diagonal singlet'' (ODS) textured by an internal orbital structure of 
compensating $d_{x^2-y^2}^{\uparrow}$ and $d_{z^2}^{\downarrow}$ spins. 
This unconventional configuration has lower energy than conventional
high-spin or low-spin alternatives.  An electronic transition is obtained at
a critical Ni-O separation $d_c^{Ni-O}=$2.03~\AA, which corresponds closely to the
observed critical value of 2.00-2.05\AA, above which Ni becomes magnetic in 
square planar NiO$_2$ compounds. 
We propose scenarios for the signature of magnetic
reconstruction in $\chi(T)$ at $T_{m}$=130 K without any Curie-Weiss background
(no moment) that invoke ordering of Ni $d^8$ moieties that are largely
this generalized Kondo singlet.
Because hole states are primarily Se $4p$ rather than O $2p$, the usual issue of
Mott insulator versus charge transfer insulator is supplanted by a character
in which electrons and holes are separated in real space.
The underlying physics of this system is modeled by a
{\it Kondo sieve} spin model (2D Kondo necklace) 
of a ``Kondo'' $d_{z^2}$ spin on each site, coupled to a $d_{x^2-y^2}$ spin that is
itself strongly coupled to neighboring like-spins within the layer. 
The observed magnetic order places BNOAS below the 
quantum critical point of the Kondo sieve model,
providing a realization of the previously unreported long-range ordered 
near-singlet weak antiferromagnetic phase. We propose electron 
doping experiments that would drive the system toward a $d^{9-\delta}$ 
configuration and possible superconductivity with similarity to
the recently reported hole doped infinite layer cuprate Ba$_2$CuO$_{3.2}$ 
that superconducts at 73K.
\end{abstract}

\maketitle

\section{Background}
Layered nickelate materials have attracted attention for some
decades primarily due to the similarities they provide to the layered cuprates that
display high temperature superconductivity.\cite{chaloupka,lanio2} 
A number of layered nickelates with
non-integer formal valences have been synthesized,
showing a great variety of magnetic and spin/charge ordering 
behavior that is now understood,\cite{botana2017,zunjie2017,botana2020} but 
no superconductivity. However, the recent discovery\cite{H.Hwang2019} and
verification\cite{Ariando2020} of superconductivity in
thin films of strained, hole-doped
NdNiO$_2$ has reinvigorated interest in layered nickelates. 
The focus has been on
$d^9$, spin-half layers, due to their similarity to the cuprates, which may be
(hole or electron) doped to produce high temperature superconducting (HTS) phases. 
The complementary route to
superconductivity, not formerly taken seriously, may arise in compounds
with a Ni $d^8$ configuration that approach the same $d^9$ regime upon electron
doping. The $d^8$ configuration brings with it the competition between 
low-spin and high-spin
configurations, which is expected to depend on strain applied to the NiO$_2$ layer
by other components of the compound as well as crystal field splittings, and 
may or may not promote HTS.

\begin{figure}[!htbp]
{\resizebox{6.6cm}{5.4cm}{\includegraphics{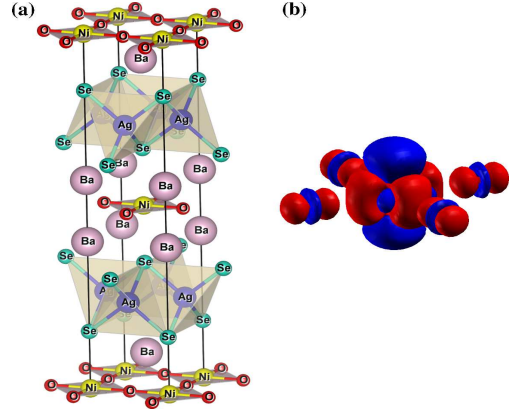}}}
\caption{(a) bct structure of \bni.
The layered square lattice of Ni ions without apical oxygen ions forms
a single ``infinite layer'' structure of the NiO$_2$ layer analogous to
the bulk layers of CaCuO$_2$ and NdNiO$_2$. Subsequent layers have a
large interlayer  `spacer layer' distance of about 10 \AA, leading to
 strong two dimensionality in many respects.
(b) Spin density plot of the spin singlet state at an isovalue of 0.015 e/\AA~
($U=7$ eV with {\sc wien2k}). Red (blue) denotes spin-up (-down) character.
This combination comprises the orbitally off-diagonal singlet state
 $d_{z^2}^{1\downarrow}$--$d_{x^2-y^2}^{1\uparrow}$.
Hybridization with the planar oxygens
is apparent.
}
\label{str}
\end{figure}

The report of synthesis of a new layered nickelate system indicates further 
anomalous properties of layered nickelates. Matsumoto and collaborators have 
reported the synthesis,\cite{matsumoto} under 7 GPa pressure at 850 $^{\circ}$C, of 
Ba$_2$NiO$_2$Ag$_2$Se$_2$ (BNOAS) containing an `infinite layer' NiO$_2$ sublattice
without apical oxygen sites but with an anomalously large Ni-O separation of 2.10 \AA. 
The susceptibility $\chi(T)$ is peculiar and unexplained.
In spite of no Curie-Weiss term in the otherwise constant $\chi(T)$, a peak
occurs at T$_m$=130 K with $30$ K full width before returning to its original value.
Field cooling below T$_m$ increases the susceptibility below T$_m$, 
and weak magnetic peaks in powder neutron diffraction\cite{matsumoto} 
at 5K have been interpreted as indicating 
$(\frac{1}{2},\frac{1}{2},0)$ antiferromagnetic (AFM) order, with best fit to
weak structure in diffraction data suggesting $S$=1 Ni spins.
This interpretation however seems inconsistent with the peculiar `spinless'
behavior of $\chi(T)$. BNOAS was discussed as insulating but no conductivity data on
the powder samples was presented.

A non-magnetic ground state is observed in other square planar $d^8$
nickelates,\cite{matsumoto,hayward} viz. BaNiO$_2$\cite{banio2}
and monovalent LaNiO$_2$ (see references in [\onlinecite{nno}])
so BNOAS is an anomalous case.
The isovalent sister strontium compound SNOAS with smaller lattice constant
is more conventional, showing Curie-Weiss
behavior above 150 K characteristic of an $S$=1 moment and strong AFM
magnetic coupling (Weiss $\theta$= --158 K). $\chi(T)$ indicates that SNOAS
undergoes some type of magnetic order around 50 K.
Matsumoto {\it et al.} ascribe the differences between the two compounds
to the different Ni-O bond lengths (that of BNOAS being unusually large),
which could tip the balance between high spin and low spin states.

Formal valence counting, confirmed by calculations presented below, proceeds as
Ba$^{2+}_2$Ni$^{2+}$O$^{2-}_2$Ag$^{1+}_2$Se$^{2-}_2$ giving Ni a $d^8$, two hole
configuration. A major issue is the relative importance of low-spin $S$=0
and high-spin $S$=1 states of the ion, and (we will propose) the nature of the $S$=0 ion.  
As mentioned, most known layered $d^8$ nickelates
are non-magnetic\cite{matsumoto,hayward} $S$=0,
so magnetic behavior (even without its strangeness) of these new
nickelates presents new physics. Building on the underlying magnetism, one can
anticipate that
electron doping, moving the Ni formal valence toward the $d^9$ configuration,
might induce superconductivity as has been found in the (Nd,Sr)NiO$_2$ system.

Our main result here is the discovery of a distinct ionic configuration, an
``off-diagonal singlet'' (ODS) with internal magnetic and orbital textures 
in spite of vanishing
magnetic moment. A secondary finding is a phase boundary at a critical
Ni-O separation of 2.03\AA~that concurs with the separation of low spin and
high spin Ni ions in square planar NiO$_2$ layered compounds.
After presenting procedures in Sec. II, 
we analyze the resulting electronic and magnetic structures
of BNOAS in Sec. III, and briefly provide some results for applied strain 
and analysis of the ODS in
Sec. IV. In Sec. V we introduce a minimal spin model for this compound 
which we feel provides a platform for accounting for the peculiar behavior 
of $\chi(T)$. Section VI provides a discussion of the various findings, with
a brief summary provided in Sec. VII.

\section{Structure and Procedures}
These compounds, which can be synthesized with several $3d$ ions, 
 have the  body-centered tetragonal (bct) structure ($I4/mmm$, \#139),
pictured in Fig.~\ref{str}(a). In the bct structure, the Ni, O, and Ag ions sit at 
high symmetry $2a$ (0,0,0), $4c$ (0,1/2,0), and $4d$ (0,1/2,1/4) sites, respectively.
The Ba and Se ions lie on $4e$ (0,0,$z$) sites with $z$=0.4136 and 0.1606, respectively. 
The NiO$_2$ layers with a large separation 
of 10 \AA~ forms a two-dimensional square lattice
without apical oxygen, as in the $d^9$ nickelates NdNiO$_2$ and LaNiO$_2$.\cite{crespin,hayward99,hayward03} 
The (AgSe)$_2$ layer provides some $\hat{z}$-axis coupling.

An intriguing feature of the
structure is that the (AgSe)$_2$ substructure is the same as that of the 
FeSe-type superconductors. The overall structure is reminiscent of many
cuprate and Fe-based superconductors: electronically active (NiO$_2$) 
layers separated by inactive ``spacer'' layers.
The strain supplied by the (AgSe)$_2$ layers results in an anomalous 
elongation of the Ni-O bond length
to 2.10 (2.05)\AA~ for BNOAS~(SNOAS), about 10 (6)\% larger than in CaCuO$_2$
and also other layered $d^8$ nickelates.
The (AgSe)$_2$ layers however are insulating and relatively inert in these compounds.

Our calculations were performed with the experimental crystal structure, using
the BNOAS lattice constants $a=$4.2095 \AA~ and $c=$19.8883 \AA, 
obtained from the neutron diffraction measurement at 5 K\cite{matsumoto},
except for a few cases where smaller planar lattice constants $a$ were 
used to probe the dependence of the Ni-O separation and identify a critical
separation.
We have compared results from two all-electron, full potential
density functional theory (DFT) codes,  
{\sc fplo-14}\cite{fplo} and {\sc wien2k}\cite{wien2k}, 
the former being an atomic orbital basis code.  The latter is a linearized
augmented plane wave code that is considered to give the most precise solutions
of the quantities that arise in DFT.
In many tests these two codes have produced results that are equivalent within
the accuracy of the underlying DFT exchange-correlation functionals. 

{\sc fplo} is a local orbital based DFT code with built-in atomic orbitals,
particularly adapted to
characterizing the electronic and magnetic structures of
transition metal compounds.\cite{fplo.basis}
In {\sc wien2k}, the basis size was determined by $R_{mt}K_{max}=7$
with augmented-plane-wave sphere radii $R_{mt}$ (in $a.u.$): Ni, 2.13;
O, 1.83; and 2.50 for
Ag, Se, and Ba. For both codes, the Brillouin zone was sampled by a dense
$k-$mesh of $20\times20\times20$ to check the energetics carefully.

We utilize the
generalized gradient approximation (GGA)\cite{gga} plus Hubbard $U$ method 
(GGA+$U$)\cite{erik}, with the double counting term of the so-called around 
mean field scheme\cite{erik,amf} as implemented in rotationally
invariant fashion in the two codes we use.
 We note that, since $U$ and $J$ are applied to 
the `atomic orbital' (which is not uniquely defined) in different ways by the
two codes, results differ somewhat. An important point of our work is that
these two codes give qualitatively similar results for which the quantitative
differences are unimportant. 

To probe correlation effects in the GGA+U calculations, 
we have at times varied $U$ in the range $1-9$ eV, always with the 
fixed Hund's spin coupling $J=0.7$ eV. 
Varying $U$ is accepted practice. There is no specific value to use 
for Ni (or any specific atom), because the large
screening of the atomic value depends on the charge state, on the local
environment, and on the conducting (or not) behavior of the system, the
latter two providing screening. Small values are more appropriate for
conducting LaNiO$_3$, larger values for highly insulating NiO.
Values of $U\approx 4-8$ eV have been applied in different 
contexts,\cite{lanio2,botana,wep98,giro05} (and see references in
[\onlinecite{nno}] for LaNiO$_2$). For BNOAS, which undergoes
a metal-insulator transition in our calculations as $U$ increases, values 
in the 5-7 eV range seem most physical. It is common 
practice to check the sensitivity of DFT+U results for sensitivity to the
value of $U$. Our results are not sensitive to $U$ within the 5-7 eV range,
or even higher.

\begin{figure}[tbp]
{\resizebox{8cm}{7.2cm}{\includegraphics{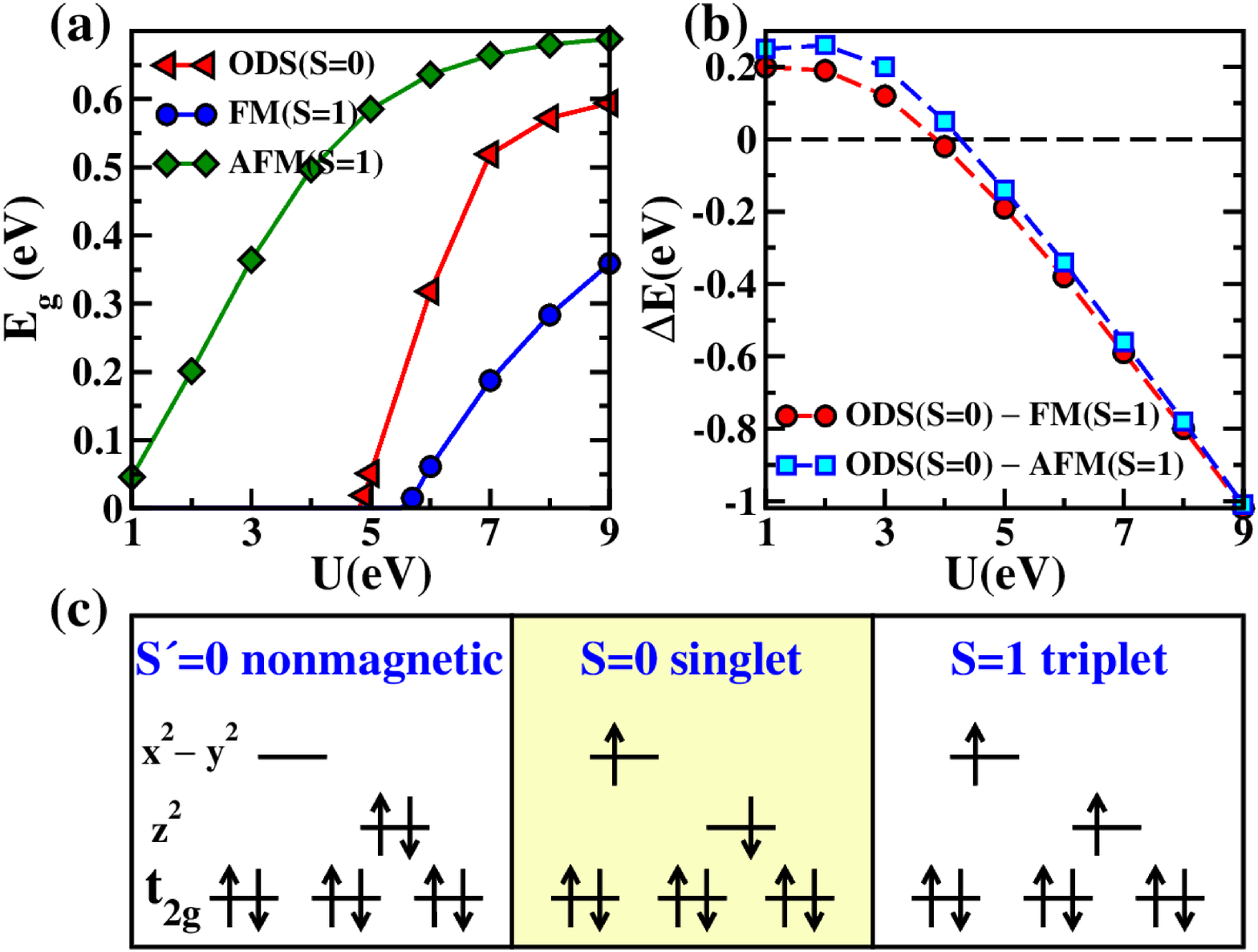}}}
\caption{Top:
Variations of (a) energy gaps and (b) relative energies among
ODS(S=0), FM(S=1), and AFM(S=1) ionic states and alignments,
as the Coulomb repulsion $U$ is varied in the 1 -- 9 eV range, from  {\sc fplo}.
Above $U_c=4$ eV, where the spin-up $d_{z^2}$ orbital is nearly emptied,
the ODS is energetically favored over the other states. The physical range
of $U$ is usually considered to be in the 5-7 eV range.
Bottom: Schematic energy level diagrams for the states of the
Ni ion; arrows indicate the spins of occupied orbitals, with the ODS in
the center panel.
}
\label{energy}
\end{figure}

\section{Electronic and Magnetic Structure}
We begin by announcing that our results
are unexpected and rather startling. The formal
$d^8$ configuration conventionally suggests that the possibilities are
(i) a high spin $S$=1 Ni ion, with holes in both the $d_{x^2-y^2}$ and
    $d_{z^2}$ minority orbitals, or
(ii) a non-magnetic low spin $S$=0 state with $d_{x^2-y^2}$ of both spins unoccupied.
We find another, (iii) an off-diagonal singlet (ODS), with holes of {\it opposite
spins} in the $d_{x^2-y^2}$ and $d_{z^2}$ orbitals
($^1B_1$ symmetry), with schematic energy level diagrams
at the GGA level presented in Fig.~\ref{energy}(c).
(The $t_{2g}$ states are nearly degenerate at the GGA level.)
The ODS is described and analyzed in Sec. III.D.

There are unusual features of the spin density, pictured in Fig.~\ref{str}(b).
Distinctive features are the following.
(1) While the net moment is exactly zero (singlet $S$=0),
there remains full spin polarization
of the two orbitals, with substantial magnetic energy benefit.
This specifies a singlet\cite{singlet} with internal texture,
which is orbital polarization.
(2) There is full spin-orbital polarization, unlike what occurs in some
insulating nickelates.
(3) This singlet is a pure spin analog of the non-magnetic Eu$^{3+}$
$f^6$ ion with $S$=3,  $L$=3, ${\cal J}=|L-S|=0$, a singlet which nevertheless has
large spin and orbital polarization.\cite{johannes,ruck2011,binh2013}
(4) It violates Hund's first rule, without apparent cause
  (such as crystal field splitting).
(5) It gains Coulomb energy by remaining maximally anisotropic (fully
orbitally polarized) within the two-orbital subspace.
Properties related to (1,2,5)
were discovered in DFT+U studies\cite{deepa2007}
of MnO under pressure. We will return to this comparison in Sec. VI.

\subsection{Energetics and Moments}
The gap values of ODS, FM, and AFM states versus $U$
are shown in Fig.~\ref{energy}(a). 
Not surprisingly, the gap opens for much smaller
$U=1$eV for AFM order due to its smaller bandwidth; for the ODS the gap opens only at
$U$=5 eV. In all cases the gap remains small for a transition metal oxide, even for
$U$=9 eV.
Next, 
we compare energies of the FM and AFM states relative to the ODS state 
in Fig.~\ref{energy}(b).
NM (trivial $S=0$ $^1A$ singlet) lacks 
spin polarization energy and can be neglected.   
Both FM and AFM ordered states have $S=1$ Ni ions with 
their Hund's rule energy gains,
with AFM being slightly favored over FM with a difference that decreases with
increasing $U$. 
The ODS state, which is {\it spin-orbital polarized} but $S$=0, 
above $U_c$=4 eV becomes
favored over both states having $S$=1 ions, with the energy
difference growing rapidly as $U$ increases. 
The Hund's exchange energy is overcome by other energy changes,
as discussed in Sec. VI.D and the Appendix.

\begin{figure}[tbp]
{\resizebox{5cm}{6.6cm}{\includegraphics{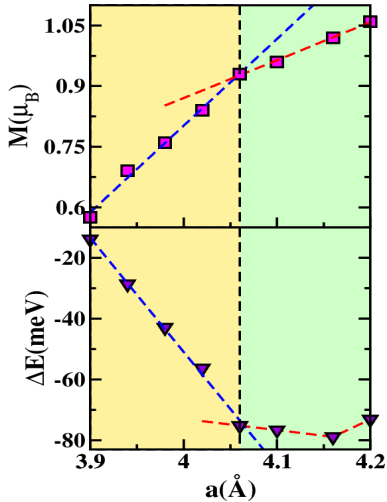}}}
\caption{Variations of (top) the Ni moment and (bottom) the
energy difference $\Delta E=E_{FM}-E_{NM}$
between FM and NM states, as the $a$ lattice constant is
varied at constant volume. From {\sc fplo} within GGA.
 Both the moment and energy difference display critical
changes as the gap closes below a Ni-O separation of
$d_c^{Ni-O}=a/2=2.03$\AA, signaling a phase change.
}
\label{moment}
\end{figure}

The values of the FM and AFM moments at GGA level are $\sim$1.1$\mu_B$, as
shown in the large lattice constant $a$ region of the upper panel 
of Fig~\ref{moment}. Including $U$, 
these grow (not shown) to around 1.9$\mu_B$,  a value that is characteristic of $S$=1
reduced somewhat by hybridization. 
To emphasize: for the ODS ions, the orbital moment is around 0.16$\mu_B$
while the net spin moment vanishes. 

\subsection{Effect of strain}
Matsumoto and coauthors\cite{matsumoto} have drawn a connection between the Ni-O
bond length and the low-spin and high-spin states
in layered $d^8$ compounds, identifying a critical separation in the
2.00-2.05\AA~range and ascribing the transition to crystal field splitting.
This critical distance led us to
study the lattice constant dependence, taking the FM state as an example.
We performed the corresponding calculations, with results shown in
Fig.~\ref{moment} using {\sc fplo} within GGA
({\sc wien2k} results are similar).
As $a$ is decreased from 4.2 \AA~to 3.9 \AA~with the cell volume fixed
and internal parameters of Ba and Se being optimized
(the observed value is $a$=4.21 \AA), an electronic transition at Ni-O is
observed.
The Ni moment initially decreases ($\Delta M/\Delta a\approx 1.6\mu_B$/\AA).
When the Ni-O separation is lowered below the critical value of $a_c/2
\equiv d^{Ni-O}_c$=2.03\AA, the rate of decrease of the Ni moment doubles.

At this same separation, the (magnetic) energy undergoes a dramatic change.
The energy increases slowly and quadratically down to
4.06 \AA. Below $a=4.06$ \AA, where Matsumoto and collaborators suggest a crossing
of magnetic and nonmagnetic Ni ions, the energy difference rises steeply 
with a slope of  --300 meV/\AA. 
This electronic transition involves a reconfiguration of how orbital occupations change
with volume.

\begin{figure}[tbp]
{\resizebox{8cm}{5.5cm}{\includegraphics{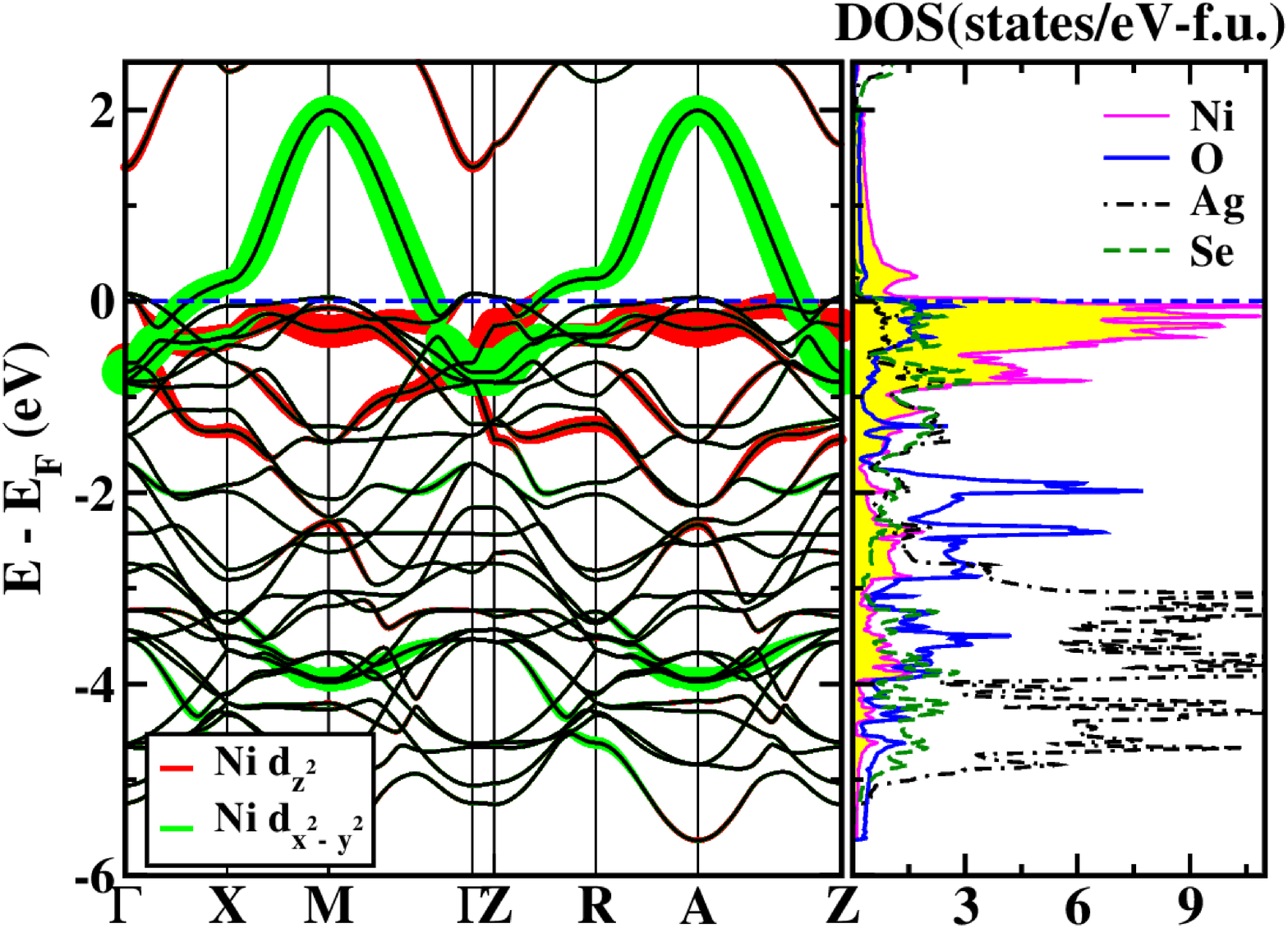}}}
\caption{Left: Reference GGA nonmagnetic band structure with highlighted fatbands
of Ni $d_{z^2}$ and $d_{x^2-y^2}$ in \bni.
Right: Corresponding atom-projected densities of states.
Most of the narrow $d_{z^2}$ band lies just below the Fermi energy $E_F$.
The mostly empty $d_{x^2-y^2}$ band has band width $\sim3$ eV,
corresponding to in-plane hopping $t\approx 0.4$ eV.
}
\label{gga}
\end{figure}

This behavior is consistent with the critical Ni-O separation proposed by
Matsumoto {\it et al}.
They observed that most NiO$_2$ compounds with only
square planar Ni (no apical oxygens) are
nonmagnetic, but if the lattice constant is pushed above 2.0\AA, the Ni
ion becomes magnetic. For example \sni, with $a$=4.094\AA, is magnetic;
BaNiO$_2$ and LaNiO$_{2.5}$ are not. The significant new aspect is that
while our Ni ion is spin-polarized, {\it i.e.} magnetism must be considered
in the electronic structure, it forms a singlet and is not
magnetic to most experimental probes.

\subsection{Electronic structure at the GGA level}
The band structure at the GGA level, which forms the basis for most beyond-GGA 
studies, is presented in Fig.~\ref{gga} with the 
atom-projected density of states (PDOS). The $d_{z^2}$ band
is degenerate with the $t_{2g}$ bands, all with 0.5-1 eV bandwidths.
The Ni $d_{x^2-y^2}$ dispersion is 3 eV (see the
$\Gamma-M$ line) with center of weight around 0.5 eV, giving an $e_g$ crystal
field splitting of 1-1.5 eV. 
The lowest conduction band at $\Gamma$ has Se $p_z$ and Ag $s$ characters;
this will be referred to as the Se-Ag band and attains importance for
electron doping, discussed below.

Dimensionality of the band structure is one central point of interest
in nickelates as well as cuprates. $k_z$ dispersion
along $\Gamma-Z$ of valence bands is small {\it except} for (1) the Ni $d_{z^2}$
band just below E$_F$, with dispersion of roughly 0.5 eV, 
accommodated by hybridization with the Se $p$ orbitals, {\it i.e.} through the electronic
polarization of the blocking layer, and (2) the Se-Ag conduction band,
which shows in-plane dispersion of 1 eV and $k_z$ dispersion from $\Gamma$ of
nearly 0.5 eV. 
The two $3d$ holes are accommodated mostly in the
$d_{x^2-y^2}$ orbitals with a minor amount in the small Fermi surfaces with
mostly $d_{z^2}$ character. These differences are central factors in how the
Coulomb repulsion $U$ will produce changes in the spectral positions. 

\begin{figure}[tbp]
{\resizebox{8.4cm}{7.0cm}{\includegraphics{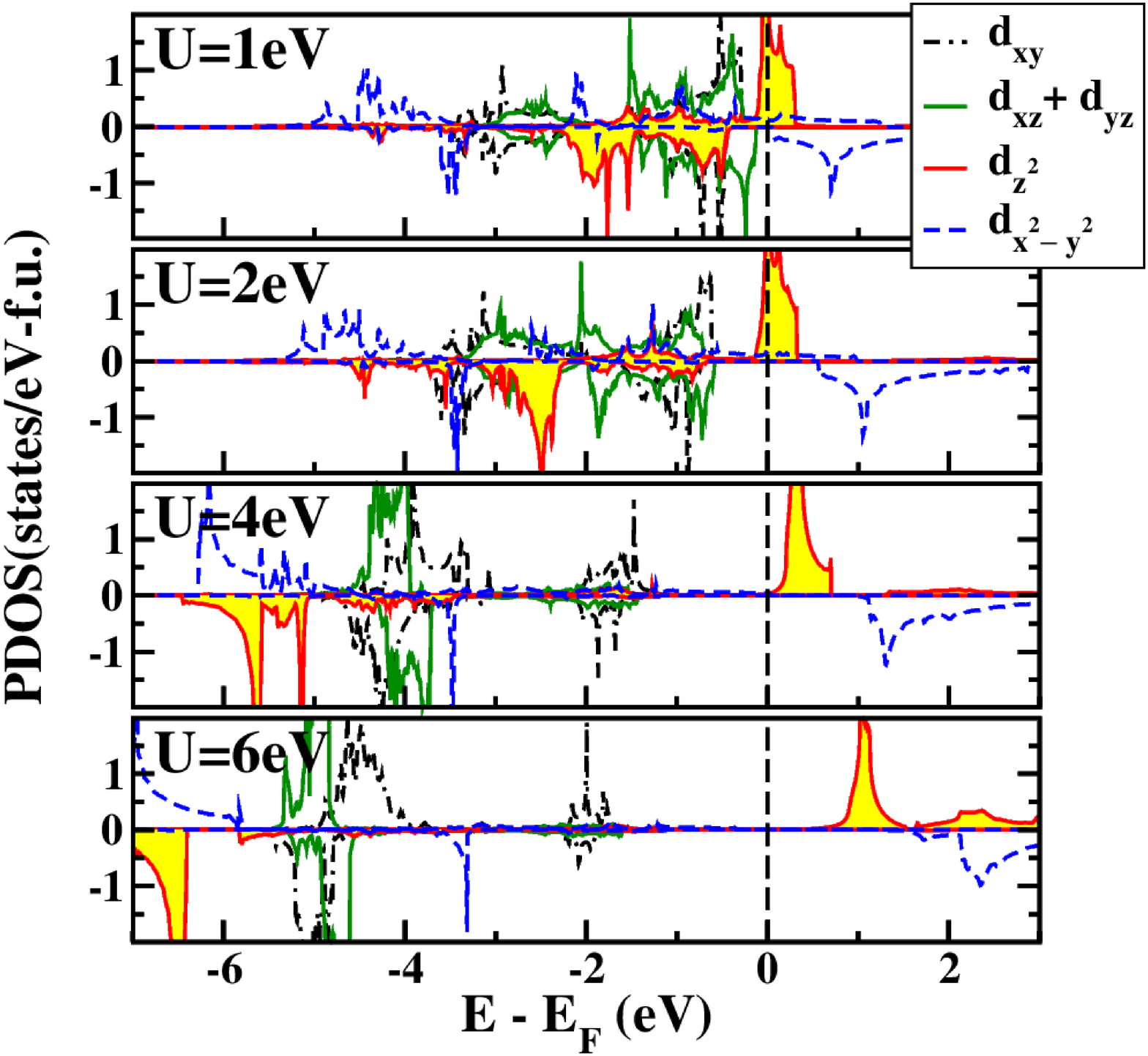}}}
{\resizebox{8.4cm}{7.0cm}{\includegraphics{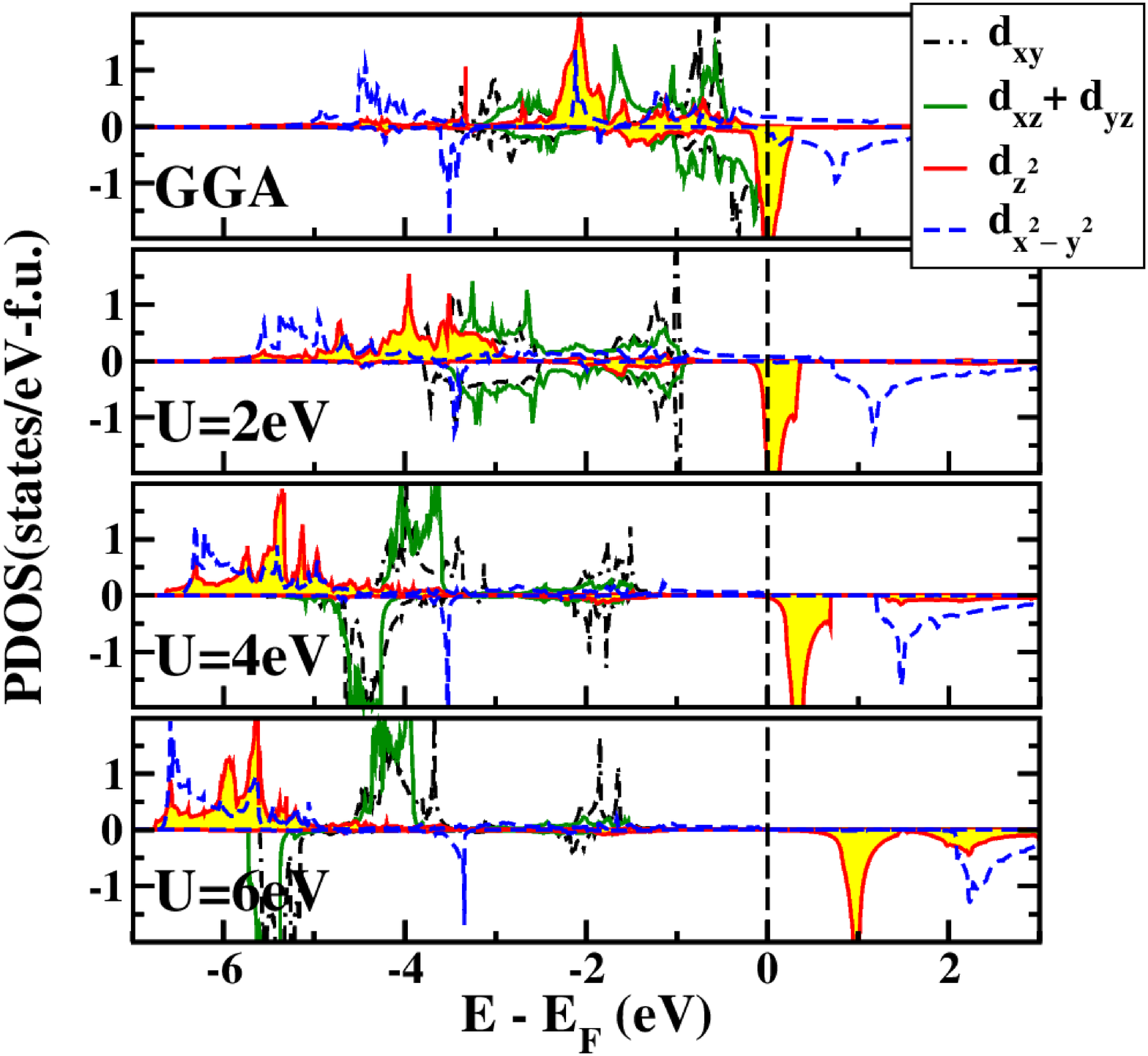}}}
\caption{Top:
the Ni $d$ orbital-projected densities of states for the ODS state
obtained from {\sc fplo}, versus strength of $U$.  Bottom: the corresponding
FM figure is shown for comparison; there is little difference except
for the direction of the unoccupied $d_{z^2}$ spin. 
Including $U$ even at the 1 eV level, the spin of the (partially) unfilled 
$d_{z^2}$ is reversed, leading to the ODS singlet state, still slightly conducting.
}
\label{udos}
\end{figure}

\begin{figure}[tbp]
{\resizebox{8.4cm}{7cm}{\includegraphics{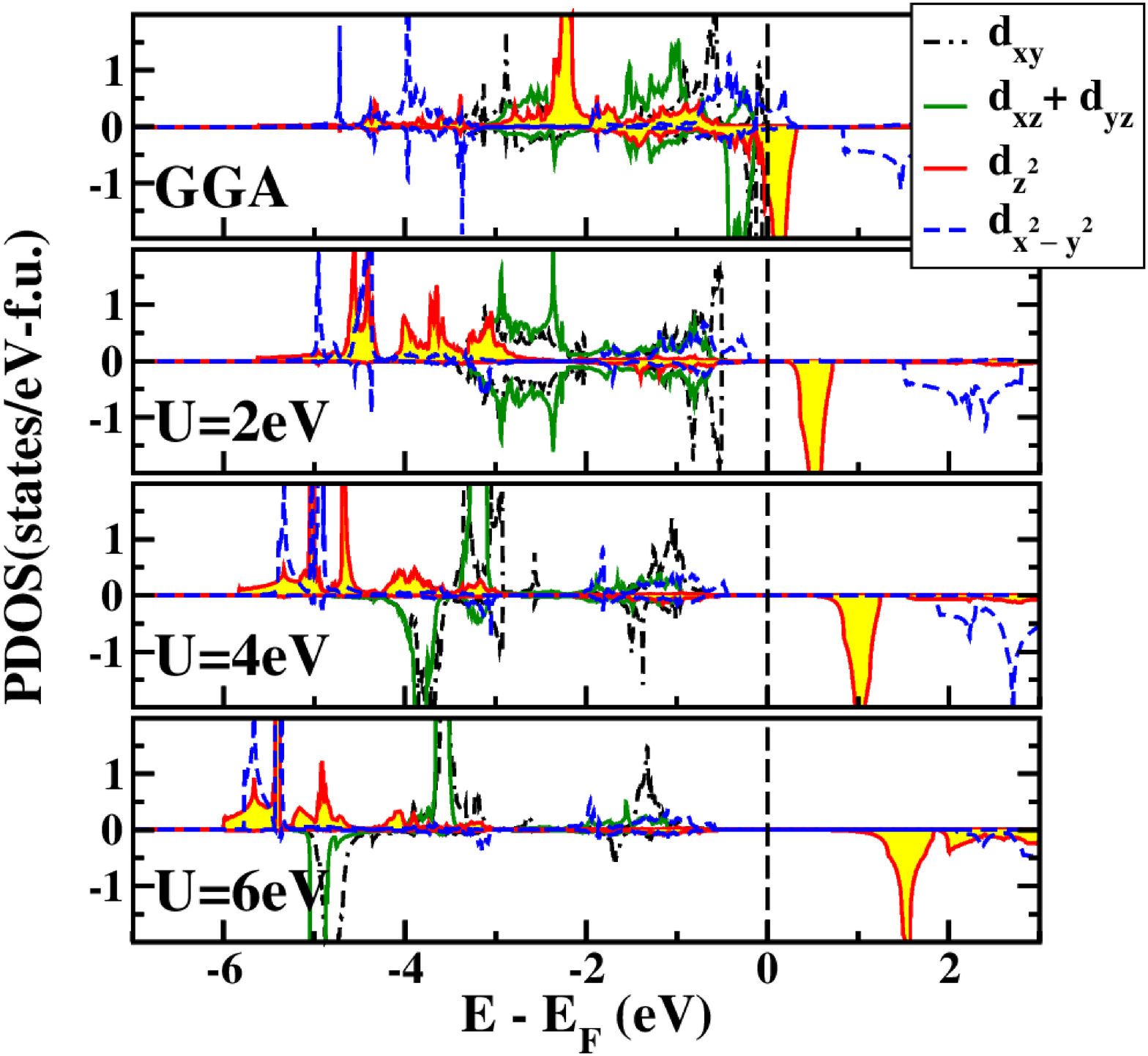}}}
{\resizebox{8.4cm}{7cm}{\includegraphics{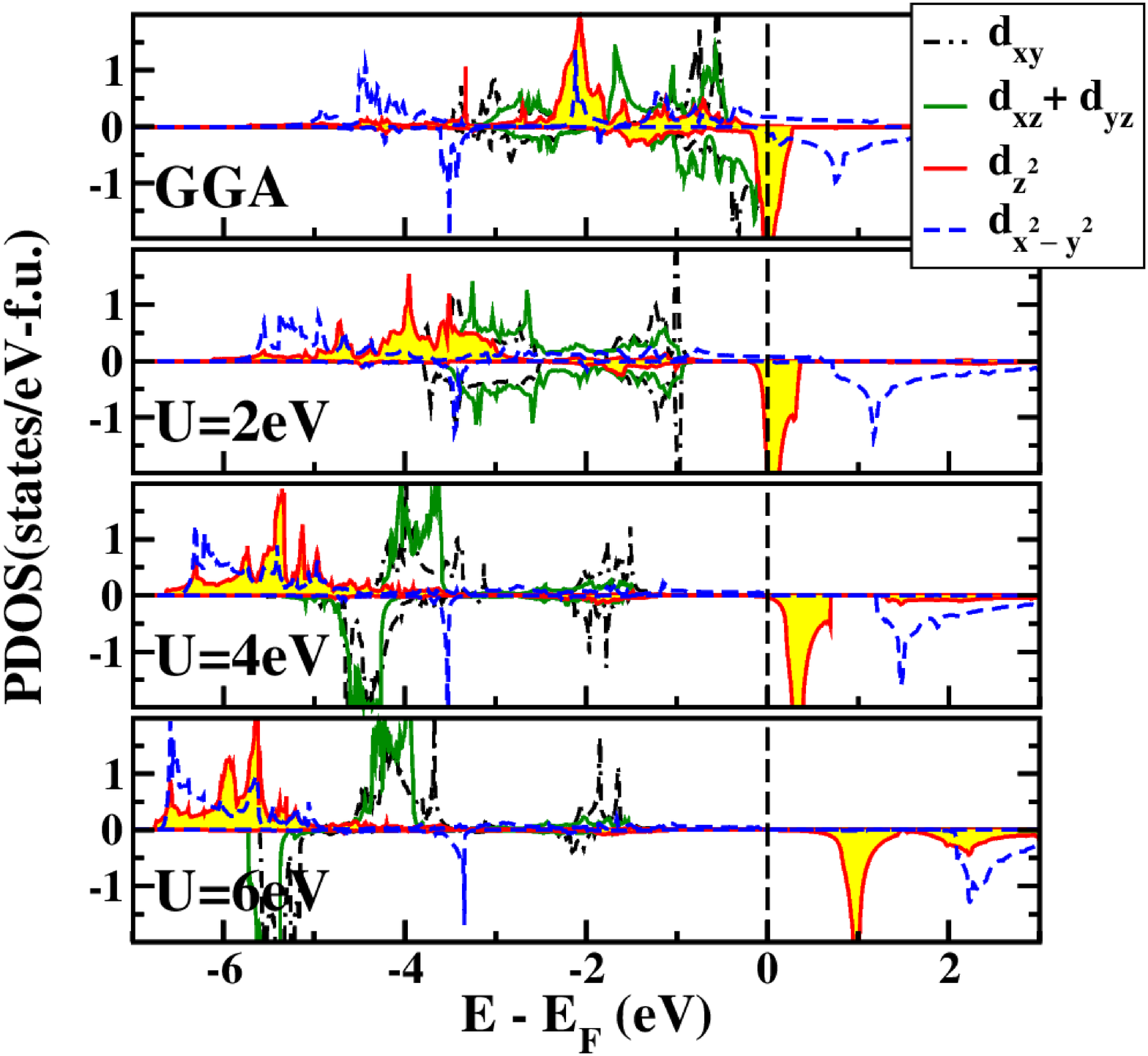}}}
\caption{Orbital-projected density of states versus $U$
for (upper panel) AFM alignment compared to
(lower panel) FM alignment, both having high spin Ni ions.
For FM alignment,
`up' is the direction of the $d_{x^2-y^2}$ spin and is plotted upward,
`down' spin is plotted downward.
For AFM, only spin-up (spin-down) indicates majority (minority) of a given Ni atom.
}
\label{AFMvsFM}
\end{figure}

\subsection{Correlation and spin polarization: the ODS}
The PDOS displayed in the top panel of Fig.~\ref{udos} reveals 
the paradigm-shifting effect of $U$.
The ODS state is obtained already for $U$=1 eV, though not yet favored energetically.  
The gap opens for $U$ in the 4-5 eV range, 
beyond which the singlet is increasingly
strongly favored (see Figs.~\ref{energy}(a,b)).  
The spin density shown in Fig.~\ref{str}(b) is strongly anisotropic
(maximally so), similar
to that obtained previously\cite{lanio2} for $d^9$ LaNiO$_2$, where 
large $U$ drove the system toward an unexpected $d^8$ configuration.
$U$=6-8 eV was judged to be too large for conducting 
LaNiO$_2$, which is observed to be conducting and would have a more
strongly screened (smaller) value of $U$.\cite{kaneko09}

Comparison with the FM progression ($S$=1) in the lower panel of the
PDOS in Fig.~\ref{udos} reveals few
differences aside from the flipped $d_{z^2}$ spin. At $U$=1eV, the occupied
(unoccupied) $d_{z^2}$ states are at -2 eV (+0.2 eV, slightly overlapping E$_F$).
There is little difference in positions and widths of most of the
occupied $d$ bands. The noticeable difference is that the occupied $d_{z^2}$
band is 1.5 eV lower for the FM case at $U$=2 eV. A close look indicates
that the $t_{2g}$ PDOS is lower for FM than for ODS.
At $U$=1 eV and above, the $d_{x^2-y^2}$ and $d_{z^2}$ `Mott gaps' 
increase proportional to $U$ with most of the separation being absorbed by
increased binding of the occupied orbital.

\begin{figure}[tbp]
{\resizebox{8cm}{5.5cm}{\includegraphics{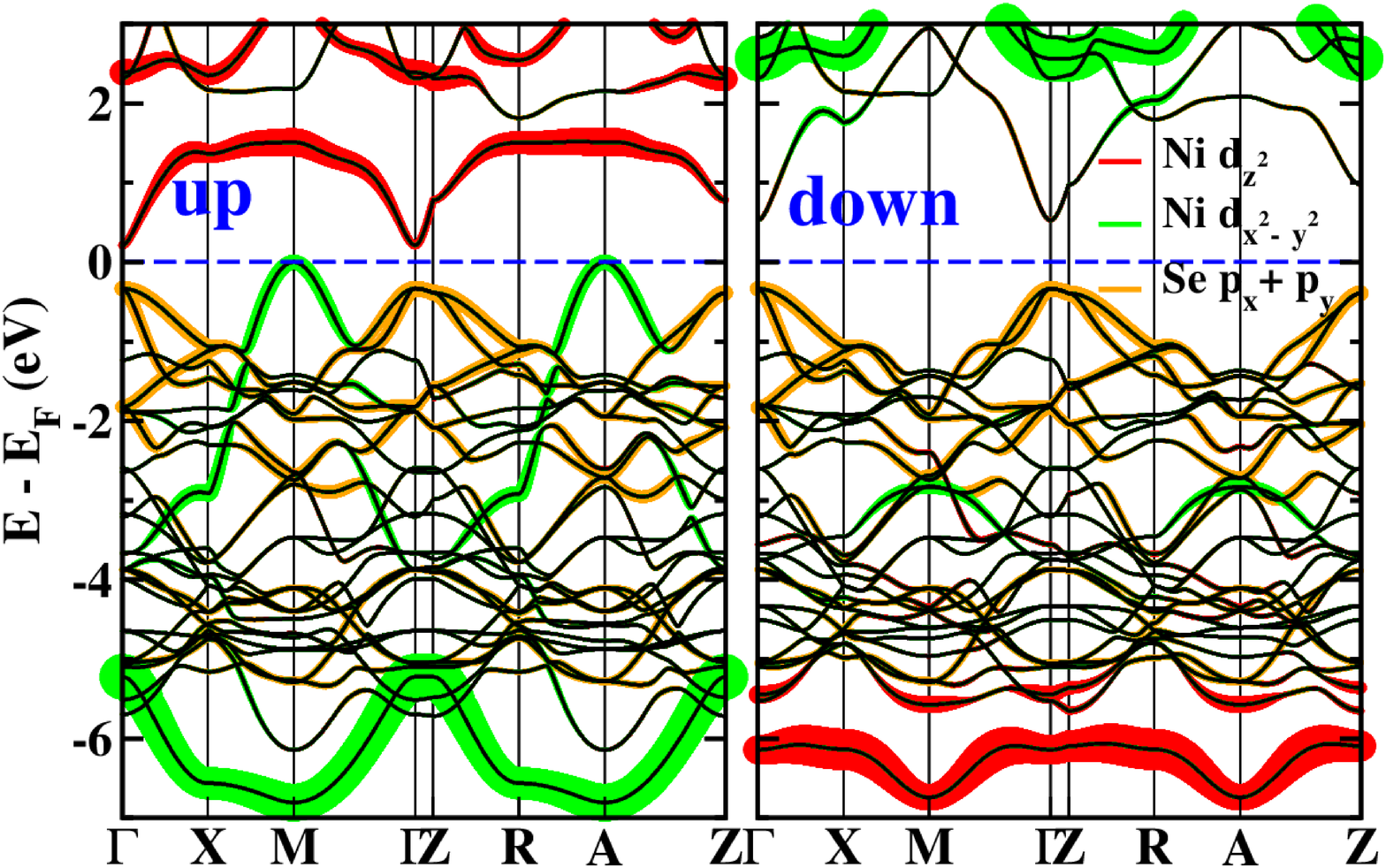}}}
\caption{Band structure of the ODS state at $U=7$ eV in GGA+U.
The fatbands of Ni $d_{z^2}$, Ni $d_{x^2-y^2}$, and Se $p_\pi$ are highlighted.
In the spin-up channel (left panel), the top of the valence bands
occurs with a Se $4p$ band at $\Gamma$ and a band at $M$ with some $d_{x^2-y^2}$
character. The valence band maximum for down spins (right panel) is a $\Gamma$-centered
band of Se $p_x + p_y$ orbitals.
}
\label{ODSbands}
\end{figure}

\subsection{High spin: AFM versus FM alignment}
In Fig.~\ref{udos}
the orbital-projected density of states (PDOS) 
of the ODS was compared to that of the FM state,
for $U$=0-6 eV.
In Fig.~\ref{AFMvsFM} we make the analogous comparison
between the AFM and FM alignments of Ni ($S$=1) spins, which we find to be instructive.
Bandwidths change, because majority states can only hop only to minority states
on nearest Ni which have different energies; thus AFM
bandwidths are narrower than FM as usual. Differences in energy position of orbital
occupations may or may not change and can be judged by this information.

Figure~\ref{AFMvsFM}
shows the progression of the Ni orbital projected DOS
as $U$ is increased from zero to 6 eV, for AFM  alignment (above) and FM
alignment (below). Each of these corresponds to a
high spin Ni ion. The $d_{z^2}$ PDOS, which is of most interest, is emphasized in yellow.
For AFM alignment, at $U$=0 holes reside in primarily in $d_{x^2-y^2}$. The $d_{z^2}$
PDOS (occupied) is spread across 3 eV. At $U$=2 eV, the gap opens giving
the full high spin state of the Ni ion.
By $U$=4 eV, the (inactive) $t_{2g}$ states stay in the -1 eV to -2 eV region.
The occupied $e_g$ majority orbitals move to lower energy and become localized,
so something nonlinear has occurred. $U$=6 eV is like 4 eV, except Mott
splitting has increased. The nonlinearity is due to the $d$ states moving
below the Se and O $p$ bands, and is not of much consequence.

For FM alignment, at $U$=0 the state is metallic but is approximating the
$S$=1 high spin state. At $U$=2 eV, tails of bands are still crossing
$E_F$ but the $S$=1 configuration is emerging.
Changes can be seen in the occupied $d$ region, as for AFM.
Increasing to $U$=4 eV, the spectrum is essentially gapped.
$t_{2g}$ states have moved to -4 eV and become localized. At $U$=6 eV the spin
splitting of $t_{2g}$ has increased, and the full $S$=1 state is attained.
There is a 1 eV difference in position between the two minority holes.
From Fig.~\ref{energy}(b) the AFM and FM energies 
are not very different.

This progression with $U$ seems about as expected: $U$ increases the separations
between occupied and unoccupied orbitals, and pushes orbital occupations toward zero
or unity. There is ``nonlinearity"
in the movement of occupied $d$ states between 2 eV and 4 eV, but this is
probably due to moving out, then being pushed out, by the O $2p$ or 
Se $4p$ bands in the
-3 eV to -1 eV region. The linear progression of the unoccupied orbitals,
widening the gap as $U$ increases, is usual.

\begin{figure}[tbp]
{\resizebox{8cm}{5.5cm}{\includegraphics{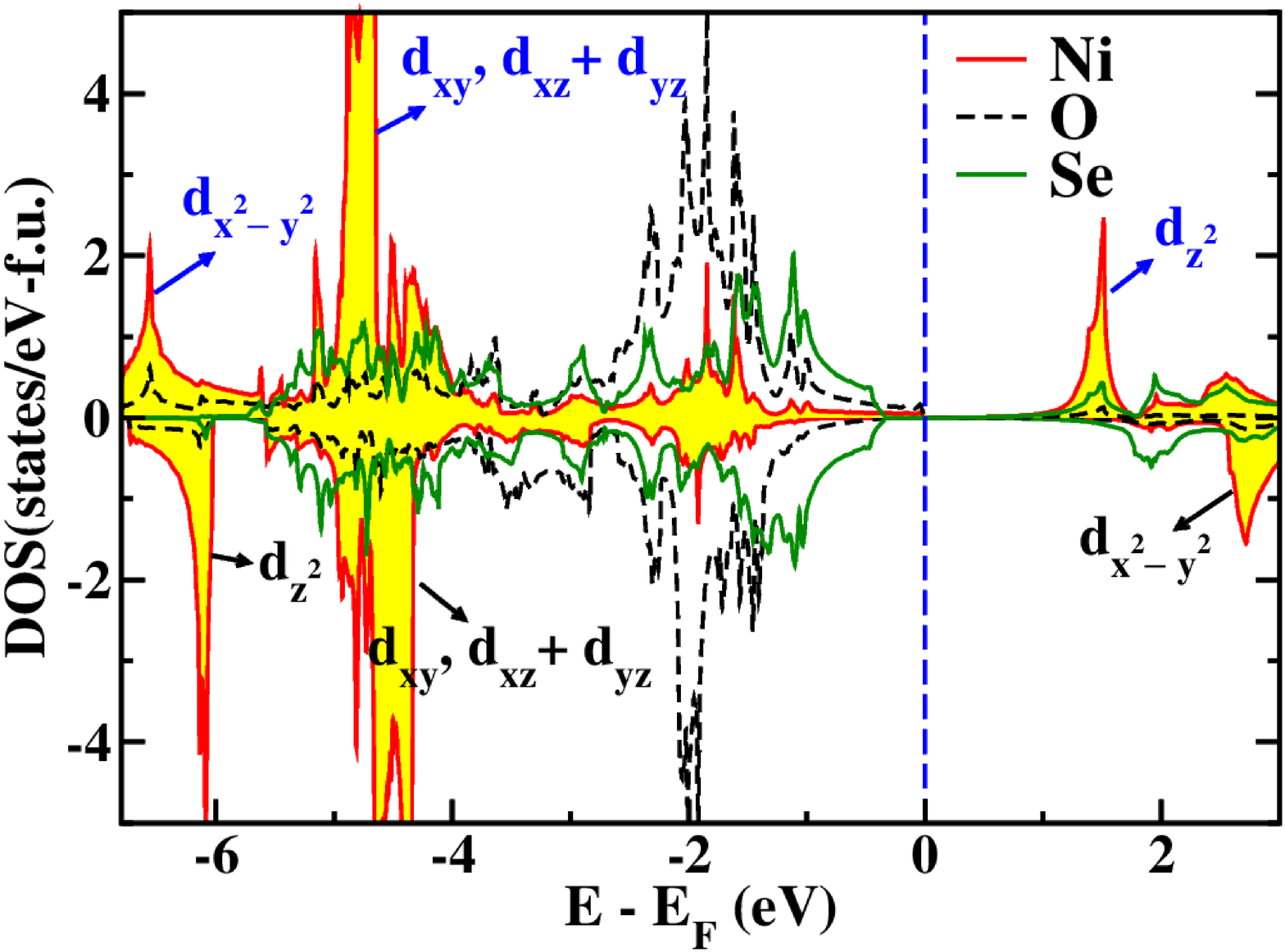}}}
\caption{Atom-projected densities of states (DOS) and Ni orbital-projected DOS
at $U=7$ eV in GGA+U for the ODS state, with
the unfilled spin-up $d_{z^2}$ and spin-down $d_{x^2-y^2}$ orbitals
being evident. The $t_{2g}$ orbitals have been lowered to -4 to -5 eV
below the gap.
}
\label{ODSdos}
\end{figure}

\section{Analysis of the ODS}
\subsection{Electronic structure}
The band structure of the ODS state is shown in Fig.~\ref{ODSbands}, with the
corresponding PDOS of relevant orbitals pictured in Fig.~\ref{ODSdos}.
Both $e_g$ orbitals are split by roughly $U$=7 eV (the Mott-Hubbard gap
between opposite spin orbitals), 
with unoccupied $d_{x^2-y^2}$
bands 1.5 eV higher in energy due to the crystal field. A substantial effect of 
$U$ is to remove the
$t_{2g}$ orbitals (both spins occupied) from just below the GGA Fermi level
to --4 to --5 eV below the gap, where they can as usual be ignored. 
This displacement serves to leave
the Se $4p$ states at the bottom of the gap at $\Gamma$, with the O $2p$
bands somewhat lower.
Note: with zero net ODS spin, we have chosen that `up' is the direction of the
$d_{x^2-y^2}$ spin, `down' is of course the opposite.

The gap opening in both spin directions at $U_c\approx$ 5 eV signals a
metal-insulator transition and ensures that the net moment vanishes,
giving the pure ODS state. This state violates Hund's first rule, signaling
that other (intra-atomic and environmental) contributions to the energy are
compensating. We propose that one factor arises from the
differences in crystal field splittings from neighboring divalent cations (Ba or Sr)
versus the tripositive ions in $d^9$ LaNiO$_2$ and NdNiO$_2$. 
There are only minor differences between {\sc FPLO} and {\sc wien2k},
what is important is that the transition to the ODS
singlet and gap opening occurs in both codes, so it is a robust feature.

The unoccupied Ni $d_{z^2}$ hole band is, in first approximation, flat at +1.3
eV. The dispersion that is seen in the Fig.~\ref{ODSbands} fatbands is due to rather
strong mixing with the Se-Ag conduction band that is dipping down at $\Gamma$.  
This hybridized band disperses
about 1 eV from $\Gamma$ to the zone edge $X$, and 0.5 eV width along $k_z$. 
The gap increases to
0.5 eV for $U$=8 eV (see Fig.~\ref{energy}), as occupancies attain 
essentially integer values.
The spin density shown in Fig.~\ref{str}(b) vividly illustrates the
spin polarization that is a combination of $d_{z^2}$ and $d_{x^2-y^2}$ contributions
of opposite spins,
with $d_{x^2-y^2}$ orbitals mixed with oxygen $p_{\sigma}$ orbitals.
We return in Sec. V to further implications of this ODS band structure.

The electron-hole asymmetry for $U>U_c$ is anomalous for a 
transition metal oxide, rendering the
usual Mott insulator versus charge transfer insulator inapplicable. Oxygen $2p$ states
do hybridize with $d_{x^2-y^2}$ but not as strongly as conventionally,
since they lie 2-4 eV below the gap. Hole states, see Fig.~\ref{ODSbands},
are instead Se $4p$ states in the down spin bands  
that are highly itinerant. O $2p$ bands are somewhat more tightly bound.

Electron states also present new behavior: electrons initially go into the
3D dispersive band at $\Gamma$ of the itinerant Se-Ag $4s$ band mixed with
up spin $d_{z^2}$ character.
There is a 0.2 eV exchange splitting, and down spin Se-Ag band does not mix
with other bands. At a still small concentration electrons begin
to occupy this spin down strongly itinerant band.
Significant doping levels will therefore produce up-carriers in the NiO$_2$
layer and down-carriers in the AgSe layer, with associated spin polarization.   
The carriers can be expected to disturb the singlet, likely enabling further
moments to emerge. A serious study of doping would require a separate study.

\subsection{Orbital occupations and charges}
\begin{figure}[tbp]
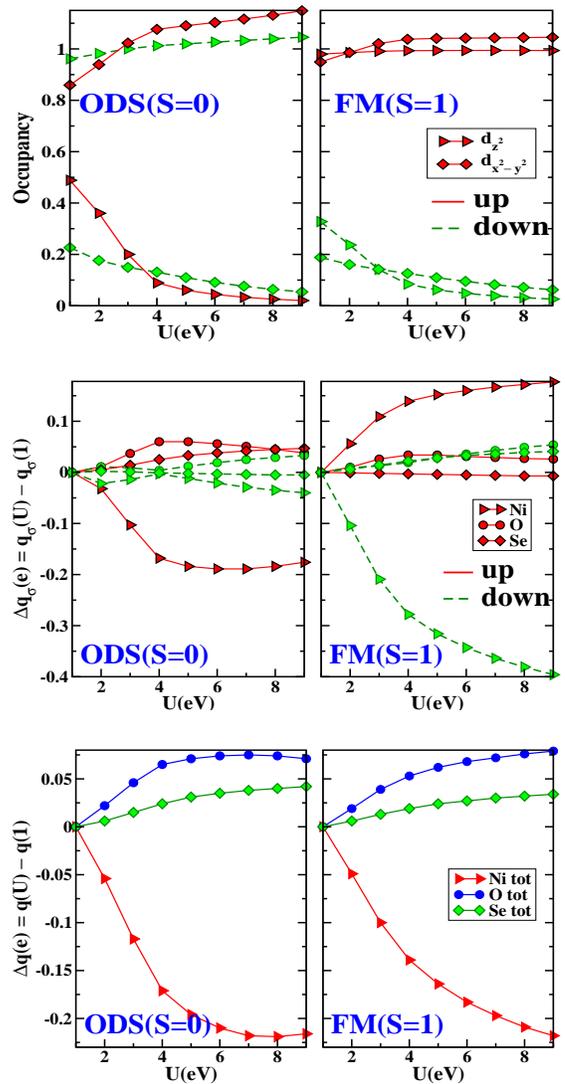

{\resizebox{7.3cm}{4.5cm}{\includegraphics{Fig9a.eps}}}
\vskip 4mm
{\resizebox{7.3cm}{4.5cm}{\includegraphics{Fig9b.eps}}}
\vskip 4mm
{\resizebox{7.3cm}{4.5cm}{\includegraphics{Fig9c.eps}}}
\caption{Variations of measures of charge of $S$=0 (singlet) and $S$=1 (FM)
states versus $U$.
Top: occupation matrix elements $n_{d_j}$ of the active Ni orbitals.
Middle (bottom): {\sc fplo} Mulliken charges of atoms as
labeled for $S$=0 ($S$=1), compared to their values at $U$=1
(because a singlet state was not obtained at $U$=0).
}
\label{3figures}
\end{figure}

Variation of spin-resolved Ni occupancies with $U$ are displayed in
Fig.~\ref{3figures}, comparing ODS with FM alignment.
The occupation numbers $n_{m}$ of the active $e_g$ orbitals $m$ (top panels)
show similar behavior especially in the physically relevant region $U>3$, except
that $d_{z^2}$ is flipped in spin in the ODS state. At the more detailed
level, the hole occupation
numbers for $S$=1 are constant for $U>$3 eV, while for the singlet $S$=0 there is a small
but continuing increase by 0.02-0.04 up to $U$=9 eV. Somewhat unexpected is that it is
$d_{x^2-y^2}$ that changes most, but is understandable because that dispersive
band crosses E$_F$ at small $U$ and $U$=5 eV is required to completely
empty it.

The atomic orbital basis of {\sc fplo} provides atomic Mulliken charges (MCs),
in which contributions to the density from overlapping orbitals of neighboring atoms is
divided equally between the atoms; the sum of MCs equals the total charge. 
The change of MCs with $U$, shown in Fig.~\ref{3figures} (middle and bottom panels),
provide a complementary view of correlation effects, since they reflect
changes in bonding (viz. relative orbital energies) as well as actual charge
rearrangement.
The atomic MCs (neutral atom values) for
 Ba,         Ni,        O,       Ag, and       Se  are
$$ 54.6 (56), 27.2 (28), 9.2 (8), 46.9 (47), 34.7 (34),$$ 
respectively.
These are quite reasonable values for MCs, and are not expected to correspond
closely to the formal charges. All do differ from the neutral atom value
in the `right direction', and by roughly half of the formal valence.

We focus on the Ni MC variation (any change on Ni must
go somewhere in the Mulliken decomposition, and that somewhere is the
atoms O and Se that hybridize with Ni).
The changes in total Ni Mulliken ionicity with increasing $U$ (lower two panels) 
are similar for ODS and FM states, saturating for
Ni $S$=0 around $U$=6 at a value of --0.2 while for $S$=1 there is less
tendency toward saturation.
The spin-resolved MC ionicities (middle panels) differ {\it qualitatively}
between FM and ODS states. For ODS $S$=0
the change in down MC is a small fraction of that of the up, which is
in the -0.18 range. For FM $S$=1 the down MC charge becomes twice as large,
while that of up is roughly half as large and of opposite sign.
At $U=1$ eV, the Ni moment is --0.18 $\mu_B$, but the net moment is zero.
With increasing $U$, the down spin charge moves to the other spin and other ions.
Above $U\approx4$ eV, the magnitude of the Ni moment vanishes,
entering a true ODS phase.
 We are not aware of nickelate or cuprate values of MCs to
compare with.

\section{Kondo Sieve spin model for BNOAS}
\subsection{The model}
The insulating ground state ODS that is discussed above
provides a system in which the
elementary excitations are spin waves. We suggest here a minimal model
of the spin system.
Let the $d_{x^2-y^2}$ and $d_{z^2}$ spin  operators be denoted by
the Pauli matrices $\vec\sigma$
and $\vec\tau$ respectively. Factors of $\frac{1}{2}$ will be folded into the
constants. The minimal spin Hamiltonian for the insulating state
contains three parameters:
\begin{eqnarray}
  H&=&  J\sum_{<i,j>}\vec\sigma_i \cdot\vec\sigma_j
       + K\sum_i \vec\sigma_i \cdot \vec \tau_i \nonumber \\
   &  &+ J_z  \sum_{[i,j]}\vec\tau_i   \cdot\vec\tau_j
  - \sum_i \vec{S_i}\cdot \vec B \nonumber \\
  &=&   H_{KS}+J_z  \sum_{[i,j]}\vec\tau_i   \cdot\vec\tau_j
          - \sum_i \vec{S_i}\cdot \vec B.
\end{eqnarray}
$J\equiv J_{\sigma}$ is the positive in-plane near-neighbor ($<i,j>$) 
superexchange constant between
$d_{x^2-y^2}~\sigma$ spins, and $K$ is the on-site spin coupling 
between $\vec\sigma$ and ``Kondo-like'' $\vec\tau$ spins, with 
contributions from Hund's coupling, Hubbard $U$ correlation, and 
environmental influences. The on-site singlet-triplet splitting is $2K$, and 
the sign of interest here (making it Kondo-like) being positive $K$. 
$J_z$ is the (spacer layer assisted) coupling between $\tau$-spin
neighbors $[i,j]$ along $\hat z$, and the total on-site spin is
$\vec {S_i}=\vec\sigma_i + \vec \tau_i$ couples to the magnetic
field $\vec B$. For discussion in this section we set units such that $J$=1.

$H_{KS}$, which we refer to as the {\it Kondo sieve} model, is an extension 
to 2D of the {\it Kondo necklace} model introduced by 
Doniach.\cite{doniach} It was derived from the 1D Kondo Hamiltonian by applying 
the Luther-Pelcher transformation\cite{luther} from 1D spinless 
itinerant particles to on-site spins. Quantum Monte Carlo calculations indicate
that the 1D Kondo necklace does not display magnetic order at any finite value
of the exchange coupling constant.\cite{scalettar1985} The extension to 2D by
Brenig\cite{brenig2006} has been known  
(somewhat contradictorily) as the 2D Kondo necklace.
This two spins per site model can also
be viewed as the large Hubbard $U$ regime of the Kondo-Hubbard model
at half filling.\cite{brenig2007}
We picture BNOAS, in first approximation, as a realization of the Kondo sieve model, then
with small and mostly frustrated interlayer coupling that may become 
important in emergence of ordered phases. 

Quite a bit is known from Brenig about the Kondo sieve model.
For $J$=0, $H_{KS}$ consists of isolated singlets; for $K$=0 it is the exhaustively
studied spin-half 2D AFM Heisenberg model, which does not order at finite T
(Mermin-Wagner theorem).  
Using stochastic series expansion methods, Brenig found that weak AFM order exists
for $H_{KS}$ at low T=0.05 up to a quantum critical point (QCP)\cite{brenig2006,brenig2007}
$K_c$=1.4.  Below $K_c$ the $\tau$-projected order
parameter is around 35\% {\it larger} than that of the 
neighbor-coupled $\sigma$ ($d_{x^2-y^2}$) spins. 

The ordering reveals that the Kondo spin promotes long range order even in 2D,
and becomes
a central influence even for a small Kondo coupling, reflecting apparent smaller
quantum fluctuations than the $\sigma$ spin. 
The uniform susceptibility decreases with increasing $K$,
vanishing at $K_c$ above which the singlets control the physics.
Site dilution of the ``Kondo'' moments, 
as by non-stoichiometry or doping,
results in structure in both $\chi(Q_{AFM},T)$ and $S(Q_{AFM},T)$ around T=0.1 or below,
depending  on concentration. The QCP vanishes into 
a crossover, {\it i.e.}
weak AFM order extends to larger $K$, for which an order-by-disorder 
mechanism was proposed.\cite{brenig2007} These results will also impact the
doping experiments proposed below.

\subsection{Relation to \bni}
Due to the body-centered stacking of NiO$_2$ layers
both the $\tau$ and also much smaller $\sigma$ couplings between AFM layers will be
frustrated. The 0.5 eV $k_z$ dispersion of the $d_{z^2}$
band imbues importance to this coupling, since it is the largest 
coupling available to enable
3D magnetic correlations and potential long-range 3D order.
An in-plane
term $\sum_{<i,j>}\vec\tau_i \cdot\vec\tau_j$ and interlayer coupling
     $\sum_{[i,j]}\vec\sigma_i \cdot\vec\sigma_j$ are symmetry-allowed,
but both should be
small.  

Of special interest, insightful for the large $K$ that our calculations
indicate, is that the on-site 
($i$) term can be diagonalized in
total spin $\vec S_i = \vec\sigma_i + \vec\tau_i$ space, with 
singlet $S_i$=0 and triplet $S_i$=1 sectors.
For the isolated ion $K$ is negative, giving the familiar high-spin Hund's
coupling. In BNOAS we find instead from total energy results that the low-spin ODS 
configuration is favored
(rather strongly) for physical values of $U$ [see Fig. \ref{energy}(b)].
The non-magnetic $^1A$ state with two holes in  the $d_{x^2-y^2}$ orbital is
highly disfavored and is not included in the Kondo sieve model.

Given that the low-spin ODS state is favored in BNOAS, in zero-th
approximation configurations
are confined to the ODS sector $S_i=0$ for all sites $i$. 
Then the second term in $H$ becomes
diagonal and the fourth
term in the Hamiltonian vanishes (and also hence the linear susceptibility
as in experiment), 
and $\vec\tau_i = \vec S_i -\vec\sigma_i$ can be
eliminated to give (up to a constant)
\begin{eqnarray}
H\rightarrow {\bar H}_{singlet}=
  J\sum_{<i,j>}\vec\sigma_i \cdot\vec\sigma_j
  + J_z  \sum_{[i,j]}\vec\sigma_i   \cdot\vec\sigma_j.
\end{eqnarray}
The in-plane superexchange $J$ is ubiquitously calculated to be positive 
and large in nickelates,
so the first term is the one that is commonly applied to the AFM insulating phase of cuprates.
Since the $\sigma$
spins are antialigned (or at least strongly correlated), the body-centering
of the BNOAS structure determines that interlayer $\vec\tau_i \cdot \vec\tau_j$ -- here
transformed to $\vec \sigma_i \cdot \vec \sigma_j$ --
coupling is frustrated.

The result from the Kondo sieve model 
without considering further symmetry breaking (such as
single ion anisotropy or structural distortion) 
is that the $\sigma$ spins are strongly
correlated in-plane  and become ordered below $K_c/J=1.4$, 
with experimental properties becoming
nonstandard for a magnetically ordered system.
The substantial Kondo coupling indicates that weak AFM should be considered
in the interpretation of magnetization data. 
Ordering across the layers due to additional interactions
may be a lower temperature possibility that is left for further work.
For comparison with experimental data, the sensitivity of the QCP to defects should be
kept in mind, since the synthesis methods of this class of compounds have
been found to allow the incorporation of impurities and produce some
minority phases.\cite{matsumoto2020}

\section{Discussion}
Our results suggest that the non-Curie-Weiss form and the peculiar structure 
in $\chi(T)$  of \bni~are couched in
formation of a Ni on-site `off-diagonal singlet' in which both $e_g$ orbitals are
singly occupied but with Kondo-like oppositely spin-directed singlets. 
This singlet forming tendency  was obtained previously\cite{lanio2} 
for infinite-layer LaNiO$_2$ at large interaction $U$ -- very strange for a
supposed $d^9$ metal --
but seemed not to be reflected in measured properties.

This singlet with intersite coupling of only one of the components leads us
to propose that a bare-bones model of \bni~is given realistically by the
Kondo sieve model (2D Kondo necklace), for which some essential characteristics
are known from quantum Monte Carlo studies.
Among the numerous theoretical viewpoints of NdNiO$_2$, the undoped parent of the
newly discovered superconducting nickelate, Zhang {\it et al.} have suggested 
that Kondo
singlets play a part in the underlying electron structure.\cite{zhang2020}
We propose that BNOAS may comprise the first realization of Kondo sieve physics,
and that it may provide important insight into layered nickelate 
physics which now includes
superconductivity when appropriately doped.\cite{H.Hwang2019,Ariando2020}
We now turn to consideration of the peculiar behavior of $\chi(T)$ reported
from experiment.

\subsection {Scenario \#1: Integrity of the Singlet}
Above the characteristic temperature T$_m=130$ K of BNOAS
the $S$=0 singlets are magnetically
inert, which naturally accounts for the lack of a Curie-Weiss term.  The internal 
structure presumably is correlated due to antiferromagnetic $d_{x^2-y^2}$ coupling, but
contributing a bit to $\chi(T)$ due to mixing of the $S$=1 triplet, 
consistent with experiment. 
In the language of the Kondo sieve model,
the ordering below T$_m$ means that $K<K_c=1.4 J$, {\it i.e.} below the QCP
where ordering occurs. 
As long range ordering of the $d_{x^2-y^2}$ internal structure
emerges at and below $T_m$, magnetic behavior emerges apparently requiring 
involvement of the $S$=1 sector, only to 
return (in zero field cooled data) to invisibility in $\chi$ in the ordered state. 
This long-range correlation and ordering of weak moments (including a small
orbital moment, which we calculate to be 0.16$\mu_B$ parallel to the
$d_{x^2-y^2}$ moment) should
be detectable, though possibly challengingly so, by neutron diffraction, by utilizing
the magnetic (spin + orbital) structure factor that contributes to the 
magnetic density correlation function. Muon spin resonance is also
a particularly sensitive method to detect magnetic order, though less so for
magnetic correlations. 
 
Probably more telling is that the susceptibility
will be distinct from that of truly non-magnetic ions, with some T-dependence
 arising from a van Vleck contribution from the proximity in
energy of the $S$=1 states. The field-cooled susceptibility below T$_m$ may arise from 
domain structure in which $S$=1 moments at the domain boundaries 
come into play.  To repeat from above: extrinsic 
phases may interfere in the data with intrinsic behavior, complicating interpretation. 

\subsection {Scenario \#2: Quenching of the Singlet}
The preceding discussion has assumed the integrity of the quantum ODS singlet. 
This is textbook behavior as long as the two spins
remain in a coherent singlet, which conventionally assumes degenerate orbitals and negligible
coupling to the environment. Neither of these conditions applies strictly to
BNOAS. At elevated temperature these ``perturbations'' may average out so the
singlet survives, but may
become more influential as temperature is lowered. Similarly to the quenching
of atomic orbital angular momentum by environmental effects (non-spherical
potential), the ODS may revert to the non-quantum pair of individual but
coupled $\vec \sigma$ and $\vec\tau$ spins (as described by DFT). 
Coupling then will strictly
antialign the individual orbital spins, which can then separately respond also to
an applied magnetic field, and have separate (and unequal) orbital contributions
and provide a distinct susceptibility though probably
still small. The magnetic transition at T$_m$ could, in this
scenario, correspond to the quenching of the quantum singlet, with consequences that
may not have been studied previously.

\subsection{Doping possibilities}
Electron doping, which would drive the Ni ion toward the $d^9$ configuration
where high temperature superconductivity (HTS) occurs in cuprates, is of primary
interest. Figure~\ref{ODSbands} indicates an unusual band structure for such a strongly
layered nickelate discussed in Sec. III.D, with the characters of the bands
bordering the gap described in Sec. IV.A. At very low doping carriers go into a 
band that disperses out of plane as much as in-plane: spheroidal Fermi surfaces. 
With increased doping the spheres will join along the short $k_z$ dimension of 
the zone, leaving roughly cylindrical Fermi surfaces. At higher dopings 
continued conduction can be expected, arising from very different spin-up and
-down contributions as discussed in Sec. IV.A: spin-up electrons in a largely
$d_{z^2}$ band, spin-down electrons in the AgSe layer.

The most obvious candidate is La$^{3+}$ substitution for Ba$^{2+}$, which gave
the initial HTS (La,Ba)$_2$CuO$_4$. Such La$\leftrightarrow$Ba substitutions 
are common, and with their electropositivites they do little other than change the
carrier concentration and to affect the volume slightly. Similarly, Ag$^{+}$
replaced by Cd$^{2+}$ might have similar effects. 

Substituting on the anion sites is the other possibility. Replacement of O$^{2-}$
by Cl$^{-}$ can achieve similar effects, with the difference that bonding with
Ni is affected to some degree. Se$^{2-}$ replacement by Br$^{-}$ is less
commonly tried (or at least achieved). Finally, one can consider 
Ni$^{2+}$ replaced by Co or Cu. These might assume the 2+ charge state, which would
give no doping but disturb the singlet character. Chemical behavior at high
pressure may determine which type of replacement is most likely to retain the
crystal structure and provide electron doping.

Hole doping seems less compelling. The states are primarily Se $4p$ in
character and essentially spin degenerate. This itinerant charge in the AgSe
layer does not change the Ni valence, and does not even screen the Ni ion,
{\it i.e.} does not change $U$ on Ni nor the gap in the NiO$_2$ layer. In
clean crystalline samples it might produce a high mobility 2D electron gas,
with its well studied properties. Having this 2D hole gas interlayered with
(ordered) ODSs would however be a novel system to study.

\subsection {Analogous materials}
Related behavior was observed in DFT+U studies of MnO under pressure,
which both experiment and theory identify as a high spin to low spin 
(but not zero spin) transition\cite{yoo2005} 
in the vicinity of an insulator-metal transition. 
In DFT+U calculations\cite{deepa2007}, the low pressure 
$S=\frac{5}{2}$ Mn ion first undergoes, under pressure,
an insulator-to-insulator transition in which two Mn $3d$ orbitals flip spin direction 
whilst each orbital remains fully
spin polarized, {\it i.e.} an $S=\frac{5}{2}\rightarrow S=\frac{1}{2}$ spin
collapse without change in orbital occupation. 
This spin collapse amounts to an even more drastic violation 
of Hund's first rule than in BNOAS that is discussed here. 
The origin of the energy gain was
traced to the persisting strong (maximal) anisotropy in spin and charge, 
and to a change in
$pd\sigma$ hybridization. In that system MnO was also in proximity to a critical 
metal-oxygen separation. 
Both of these characteristics (anisotropy and a critical bond length) are also apparent 
in BNOAS.

A prime interest in BNOAS will be how it may inform the electronic structure 
and magnetic behavior in superconducting (Nd,Sr)NiO$_2$. As mentioned, the
ODS tendency was seen fifteen years ago in LaNiO$_2$, which is commonly
used as a stand-in for NdNiO$_2$ in correlated electron calculations.  
It should be mentioned that others have not reported obtaining this singlet
tendency in LaNiO$_2$. 
However, in BNOAS we obtain it with two different codes and GGA+U implementations. 
We obtain the
magnetic states others report but they have higher energies (for physical values
of $U$). There are still
things to be learned about the peculiar behavior of the $d_{z^2}$ orbital in infinite
layer nickelates. It is of considerable interest to try to electron-dope
BNOAS toward the $d^{9-\delta}$ regime where superconductivity emerges in
(Nd,Sr)NiO$_2$. 

Another related system is Ba$_2$CuO$_{3.2}$, which superconducts at 73K\cite{pnas}
after having been synthesized at 18 GPa and 1,000C.
Although not described as such, this compound is an infinite layer cuprate
that is hole-doped with 0.2 apical oxygen atoms per copper, making it formally
somewhat overdoped. The (average) apical Cu-O distance is small at 1.86\AA,
whereas the in-plane Cu-O separation is 2.00\AA, larger than other superconducting
cuprates. The interpretation of spectral data suggested that the crystal field
has exchanged the positions of the $d_{z^2}$ and $d_{x^2-y^2}$ levels, which
would provide a new paradigm for cuprate superconductivity. The relationship
of this material to BNOAS suggests the usefulness of further study of both
systems.

While BNOAS is not simple to synthesize due to the need of high pressure, it is
nevertheless available (after synthesis) to experimental probes that are 
not available to MnO at 100 GPa pressure.
We encourage further synthesis and study of BNOAS by thermodynamic and spectroscopic
(electron, neutron, and muon) probes to illuminate the peculiar magnetic
behavior observed in this infinite layer nickelate.

\section{Summary}
Our density functional theory with correlation effects study of \bni~has uncovered
the likelihood that an off-diagonal singlet arises in the subspace of the two
$e_g$ holes (or electrons) in its $d^8$ ionic configuration. This ``low spin''
singlet however has orbital polarization, {\it i.e.} orbital texture, with one
orbital's spin being coupled to neighbors while the other is not. Moreover, a
critical Ni-O separation of 2.03\AA~has been obtained, which seems to be associated 
with the distinction between low spin and high spin Ni $d^8$ ions in square planar
coordination, occurring at the same Ni-O distance. Two scenarios have been 
presented, either of which might account for the peculiar measured susceptibility:
a magnetic transition at T$_m$=130K but non-magnetic behavior above T$_m$.

\vskip 2mm
\section{Acknowledgments}
We acknowledge assistance from Young-Joon Song and Mi-Young Choi in the
early stages of  this work. 
The spin model aspects of this paper have
benefited from feedback and references from R. T. Scalettar and comments
from R. R. P. Singh.
We appreciate extensive comments on the manuscript from M. D. Johannes
and thank A. B. Shick 
for the use of his code that calculates the anisotropic $U$ and $J$
matrices reported in the appendix.
H.S.J and K.W.L were supported by National Research Foundation of Korea 
Grant No. NRF2019R1A2C1009588.
W.E.P. was supported by National Science Foundation Grant DMR 1607139.

\appendix
\section{Anisotropy of $U_{m,m'}$ and $J_{m.m'}$}
The $U$ and $J$ matrix elements introduce orbital-occupation differences into the
potentials and energies of the DFT+U results that are then included
self-consistently. We provide these matrix elements for
study of some of these differences that arise.
The importance of on-site anisotropy has been noted previously but several groups.
The study of the `flipped spin' phenomenon in MnO under pressure by
Kasinathan {\it et al.}\cite{deepa2007}, mentioned in the main text, revealed that
the anisotropy of Coulomb repulsion and exchange parameters is deeply involved
in the energetics causing that flipped spin from $S=\frac{5}{2}$ to $S=\frac{1}{2}$
while all $3d$ orbitals of Mn remain fully spin polarized.

The situation we find in BNOAS
is strongly analogous, so we provide below the matrix elements that enter the
DFT+U calculations.
We provide the matrix elements for $U$=5 eV, $J$=0.7 eV.
The Slater integrals (eV) are
F$_0$=5.00,  F$_2$=6.03,  F$_4$=3.77,
and the corresponding orbital-resolved matrix elements that occur in the DFT+U
functional (in eV) are given just below. This connection between the Slater
integral and $U$ and $J$ is important to understand. 

Features to notice are (1) $U_{mm}=J_{mm}$: self-exchange equals
self-interaction, which the functional attempts to account for, and (2)
the anisotropies in $U_{mm'}$ and $J_{mm'}$ are both proportional to $J$.
$J$ describes both the anisotropy of the Coulomb repulsion and the
anisotropy of the exchange interaction. Setting $J$ equal to zero results
in a $U$ matrix with all elements equal to $U$, and of course a diagonal
$J$ matrix with diagonal elements equal to $U$ -- no anisotropy whatsoever.

The matrices are given below in the complex $\ell,m$ representation of $3d$
orbitals. In this representation the matrices are symmetric around both
diagonals.

\begin{table}[ht]
The $U$ matrix:
\begin{center}
\begin{tabular}{c|ccccc}
 \multicolumn{1}{c|}{$m_l$}&
 \multicolumn{1}{c}{-2}&
 \multicolumn{1}{c}{-1}&
 \multicolumn{1}{c}{0}&
 \multicolumn{1}{c}{+1}&
 \multicolumn{1}{c}{+2}
   \\ \hline
-2~&~5.501&~4.720&~4.559&~4.720&~5.501 \\
-1~&~4.720&~5.260&~5.041&~5.260&~4.720 \\
0~&~4.559&~5.041&~5.800&~5.041&~4.559  \\
+1~&~4.720&~5.260&~5.041&~5.260&~4.710 \\
+2~&~5.501&~4.720&~4.559&~4.720&~5.501 \\
  \hline
\end{tabular}
\end{center}
\label{matrixtable1}
\end{table}
\begin{table}[ht]
The $J$ matrix:
\begin{center}
\begin{tabular}{c|ccccc}
 \multicolumn{1}{c|}{$m_l$}&
 \multicolumn{1}{c}{-2}&
 \multicolumn{1}{c}{-1}&
 \multicolumn{1}{c}{0}&
 \multicolumn{1}{c}{+1}&
 \multicolumn{1}{c}{+2}
   \\ \hline
-2~&~5.501 & 0.781 & 0.621 & 0.299 & 0.598\\
-1~&~0.781 & 5.260 & 0.380 & 1.080 & 0.299\\
 0~&~0.621 & 0.380 & 5.800 & 0.380 & 0.621\\
+1~&~0.299 & 1.080 & 0.380 & 5.260 & 0.781\\
+2~&~0.598 & 0.299 & 0.621 & 0.781 & 5.501\\
  \hline
\end{tabular}
\end{center}
\label{matrixtable2}
\end{table}

\vskip 2mm
Suppose all $3d$ orbitals are fully occupied or empty $n_{d_i}$= 1 or 0.
FM $S$=1 and singlet $S$=0 have same two $e_g$ orbitals
occupied (and same two unoccupied, with opposite spin).
The +$U$ term is spin-blind, so no difference. In the $J$ matrix,
the parallel-spin $S$=1 term has $J_{x^2-y^2,z^2}$, compared to
zero for $S$=0 ($J$ term is parallel spin only). Actually, one
should really count all the pairs of occupied states instead of unoccupied
states. Anyway, term this
favors the high-spin, not the singlet, state (by 0.62 eV), as
simple Hund's rule suggests but gains more from the off-diagonal repulsion $U$. 
Kasinathan {\it et al.} provide an analysis
of how the Hund's exchange energy can be overcome, with the above
anisotropy being a central player.


\begin{thebibliography}{10}
\bibitem{fplo.basis} 
The orbitals retained in the band problem are Ba $5s5p6s(7s5d6p)$, 
Ni $3s3p3d4s(4p4d5s)$, O $2s2p(3s3p3d)$, 
Ag $4s4p4d5s(5p5d6s)$, and Se $3s3p3d4s4p(4d5s5p)$. 
The orbitals in parentheses indicate semicore or polarization orbitals.
\bibitem{singlet}Phrasing more properly, this is the band representation
of a spin polarized but zero net spin state. DFT does not deal 
with several electron wavefunctions,
thus it has no representation of a quantum singlet.
\end{thebibliography}
\end{document}